# A District-level Ensemble Model to Enhance Dengue Prediction and Control for the Mekong Delta Region of Vietnam

Wala Draidi Areed, Thi Thanh Thao Nguyen, Kien Quoc Do, Thinh Nguyen, Vinh Bui, Elisabeth Nelson, Joshua L. Warren, Quang-Van Doan, Nam Vu Sinh, Nicholas Osborne, Russell Richards, Nu Quy Linh Tran, Hong Le, Tuan Pham, Trinh Manh Hung, Son Nghiem, Hai Phung, Cordia Chu, Robert Dubrow, Daniel M. Weinberger,*† Dung Phung*

*These authors contributed equally

†Correspondence to daniel.weinberger@yale.edu

**Abstract**

Background

The Mekong Delta Region (MDR) of Vietnam is increasingly vulnerable to severe dengue outbreaks due to urbanization, globalization, and climate change. Early warning systems are essential for effective outbreak mitigation. This study develops a probabilistic forecasting model to predict dengue incidence and outbreaks, enabling proactive interventions to reduce dengue impacts.

Methods

Prediction models with 1–3-month lead times were developed using meteorological, sociodemographic, preventive, and epidemiological data. Seventy-two models were evaluated, and top performers from spatiotemporal models, supervised PCA, and semi-mechanistic hhh4 frameworks were combined into an ensemble. Data from 2004-2011 supported model development, 2012-2016 for cross-validation, and 2017-2022 for evaluation. Performance was assessed using Brier Score, Continuous Ranked Probability Score (CRPS), bias, and diffuseness.

Findings

The ensemble model, integrating five individual models, forecasted dengue incidence up to three months ahead with performance varying by horizon, geography, and seasonality. Using the 95th percentile of the historical distribution as the epidemic threshold, it achieved 69% accuracy at a 3-month horizon during evaluation, outperforming a reference model's 58%. However, it struggled in years with atypical seasonality, such as 2019 and 2022, potentially impacted by COVID-19 disruptions.

Interpretation

The model offers critical lead time for health systems to allocate resources, plan interventions, and engage communities in dengue prevention and control.

Funding

This research was fully funded by the Wellcome Trust [Grant number 226474/Z/22/Z]. A CC-BY public copyright license has been applied to any author-accepted manuscript version arising from this submission.

**Introduction**

Dengue, a vector-borne disease transmitted by *Aedes aegypti* and *Aedes albopictus* mosquitoes, poses a significant global public health threat, especially in tropical and subtropical regions. Dengue is now endemic in more than 125 countries[1]. Globally, its incidence increased tenfold between 2000 and 2019 to more than 5 million reported cases, with 6·5 million cases reported in 2023[2]. During the first nine months of 2024, an unprecedented dengue upsurge occurred, with more than 13 million reported cases[3]. The reported rate of dengue likely represents an underestimate due to testing patterns, and it has been estimated that there were actually 59 million cases in 2021[4]. The global spread of dengue has been driven by urbanization, globalization with its associated increase in human mobility, and climate change [5–8]. The global economic cost of dengue was estimated to be US$8·9 billion in 2013[9].

Most dengue cases are subclinical or present as a non-specific febrile illness with flu-like symptoms[10, 11]. However, 2-5% of symptomatic cases progress to severe dengue, a life-threatening condition associated with bleeding and shock[10]. Treatment is supportive, including oral or fluid replacement depending on illness severity [10]. Vector control and preventing mosquito bites historically have been the main approaches to dengue prevention and control[10, 11]. Novel vector control methods, including release of sterile, genetically modified, or *Wolbachia* infected mosquitos, have been developed but are not yet in widespread use [10, 11]. Three different tetravalent attenuated virus vaccines, targeting the four serotype of dengue, have shown efficacy in randomized controlled trials [11–13], and two of these have been licensed [11, 12]. However, vaccines are not yet widely used due to a number of challenges[10, 11].

Vietnam, where dengue is endemic, is a high-incidence country, although its annual incidence tends to be cyclical. [13] In 2023, the number of reported dengue cases in Vietnam was 369,000, the second highest in the world [1], with an estimated 1·1 million actual cases in 2021 when adjusting for underreporting [14]. The Mekong Delta Region (MDR) of Vietnam has been particularly affected, with 40-50% of reported cases in Vietnam occurring in the MDR during 2000-2016 [13]. Drivers of dengue incidence in Vietnam include urbanization, ambient temperature, and hydrometeorological variables [15]. The economic cost of dengue in Vietnam was estimated to be US$94 million in 2013 [9]. Vietnam's current dengue prevention and control strategies, as outlined in the Ministry of Health's Decision No. 3711/QD-BYT dated 19/9/2014, rely predominantly on vector control, such as entomological monitoring, community mobilisation to eliminate breeding sites, and insecticide spraying [16]. However, these interventions are often reactive, only implemented after cases have been reported, which could be too late to pre-emptively reduce the impact of dengue outbreaks

A great deal of work has been devoted to developing models to predict dengue incidence and outbreaks [17] – such models could be used to develop early warning systems to enable proactive interventions. Investigators recently developed an ensemble of probabilistic dengue models to predict dengue incidence one to six months in advance at the provincial level -- the largest geographical and administrative unit -- in Vietnam [18]. While this approach offers a valuable tool for improving dengue surveillance and planning preventive interventions, a critical limitation is that the predictions are made at the provincial level, which may not be sufficient for localised action. In Vietnam, dengue prevention and control measures are implemented at the district level, with a typical province in the MDR composing from seven to 15 districts, making it necessary to refine models to provide district-level predictions that align more closely with the administrative and operational realities of dengue control programs.

This study aims to evaluate the performance of a range of complementary probabilistic models for forecasting the incidence of dengue 1-3 months in advance at the district level in the MDR, incorporating meteorological, epidemiological, sociodemographic, and preventive intervention data. To enhance the accuracy and robustness of forecasts, we integrated five models into an ensemble. In future work, we aim to incorporate our ensemble model into user-friendly software for an early warning system to assist districts in proactively implementing interventions to mitigate dengue outbreaks.

METHODS

*Overview of approach*

The goal for these analyses was to generate and evaluate probabilistic forecasts for dengue cases and outbreaks at the district level, one to three months in the future and to combine the best of these models into an ensemble model. We evaluated three distinct modelling approaches– hierarchical spatio-temporal Bayesian models, semi-mechanistic spatiotemporal (hhh4) models, and a machine learning supervised principal components analysis regression -- each with various sets of variables. The data were split into three periods: the initial set covering the years 2004-2011, the set from 2012-2016 for time series cross validation, and the evaluation set from 2017-2022 to assess the performance of the ensemble model out-of-sample (Fig S1).

*Data sources*

*Epidemiologic data*

We obtained detailed monthly dengue data for each district (n = 134) in the MDR from the Vietnamese Ministry of Health, covering September 2004 to December 2022. Dengue cases, reported within 24 hours via the electronic communicable disease system, are recorded as either suspected or confirmed. Surveillance quality is maintained by comparing reported data with hospital records. Most cases were suspected, with limited confirmation through diagnostic tests [19].

Administrative boundaries changed during the study, with the latest adjustment in 2015. To maintain consistency, districts with boundary changes during the study period were merged, reducing the dataset to 114 districts. Two districts without neighbouring boundaries were excluded, resulting in an analysis of 112 districts.

*Weather data*

Weather data were sourced from two datasets: the ERA5-Land reanalysis database and ground-based *in situ* weather stations. These datasets were used separately as input for prediction models.

The ERA5-Land reanalysis dataset, developed by the European Centre for Medium-Range Weather Forecasts, provides high-resolution (~10 km) global coverage tailored for land-specific meteorological and hydrological data, particularly beneficial for areas with varied landscapes and climates (Muñoz Sabater, 2019). Daily minimum, maximum, and average temperature, relative humidity, and cumulative rainfall were collected from September 2004 to December 2022. To ensure compatibility, ERA5-Land data were clipped to align with the administrative boundaries of each district within the MDR. For smaller districts or those with boundaries not fully overlapping an ERA5 grid cell, nearest-neighbour interpolation was applied, assigning the closest ERA5 grid cell values to such districts.

The ground-based data consisted of 18 weather stations in and around MDR. These data are sourced from the Vietnam Meteorological and Hydrological Agency, which record daily minimum, maximum, and average temperature, relative humidity, and cumulative rainfall. Due to the limited number of stations relative to district-level needs, Kriging interpolation was applied to estimate district-specific climate conditions. Kriging, a geostatistical technique, is widely used in climate science to predict values at unsampled locations by accounting for spatial correlations in climate variables. This approach uses semi-variograms to model spatial variance and correlation, producing accurate predictions and measures of uncertainty, which is particularly advantageous in regions with sparse data [20].

Consequently, two separate weather datasets (i.e., reanalysis and weather stations), including daily time series for each variable of interest, were generated for each district over the specified period. For use as covariates in the dengue prediction models, daily temperature and humidity data were then aggregated into average monthly values, and rainfall data were aggregated into cumulative monthly values.

We obtained the weather variables from two sources, ERA5 and local weather stations, and tested the performance of the models for each. Since both of these two sources provided similar results, we reported in the paper the results from ERA5 due to its simplicity and ease of access.

Other data, including sociodemographic factors (e.g., urbanization, population density) and preventive measures/entomologic indices (e.g., spraying, breeding site elimination campaigns, training, Breteau index), were tested but did not significantly enhance predictive performance and were excluded from the final model. Further explanation for these variables can be found in Tables S1 and S2.

*Outbreak definitions*

Defining epidemic thresholds for dengue is not straightforward, and there is little agreement about what constitutes an appropriate method for setting thresholds [21]. We therefore evaluated several possible definitions of outbreaks.

1) Mean + 2 standard deviations: calculated using the mean and standard deviation from the same month for each district over the previous five years. If an outbreak occurred during this period, it was excluded, and data from an earlier year were used to improve threshold sensitivity [21]. Months with number of cases above this threshold were considered to have an outbreak.

2) 95th percentile: uses data from the same month and the same district across all available years to determine the 95% threshold. Months with number of cases above this threshold were considered to have an outbreak.

3) Poisson: a generalized linear model with Poisson likelihood predicts epidemics based on monthly data for each district. Simulations from the model account for parameter uncertainty and observation uncertainty, and the 97.5th percentile of these simulations serves as the epidemic threshold. Months with observed cases above this threshold signal an outbreak.

4) Fixed threshold based on incidence rates: the monthly threshold was set at a fixed level: 20, 50, 100, 150, 200, or 300 cases/100,000 population.

*Models*

We evaluated three types of modelling approaches:

1) Hierarchical Bayesian spatiotemporal models. These models tested lags of different covariates including meteorological variables (e.g., temperature, precipitation), sociodemographic variables (e.g., urbanization, income), and preventive measures/entomologic indices (e.g., active spray, communications or training, Breteau index). The models also included 1-3 lags of observed monthly dengue cases and spatiotemporal random effects with different structures to account for different sources of correlation and ensure accurate statistical inference (i.e., autoregressive correlated terms modelled independently across each district, spatial random effects, spatiotemporal random effects). This work builds upon the approach developed by Colón-González[18] for monthly data at the provincial level in Vietnam, with a couple of simplifications. First, we used weather variables and the observed monthly dengue case variable lagged by the amount of the maximum forecast horizon (e.g., for models predicting up to 3 months in the future, the minimum lag of observed cases and weather variables was 3 months)—this avoided the need to recursively use forecasts of the weather variables as predictors in the model. There are numerous possible combinations of covariates and random effects structures, resulting in the evaluation of 60 models (see Table S3). To ensure a thorough exploration of potential covariates, we tested at least one representative covariate from each covariate type (meteorological, sociodemographic, preventive measures/entomologic indicators, monthly dengue cases) for the different sets of covariates. To address high correlations between some variables, we avoided including highly correlated variables in the same model (Table S3).

2) hhh4 models. These semi-mechanistic spatiotemporal models account for variations in baseline incidence, autocorrelation, and spatial spread. Each of these three components are modelled as a function of covariates. The endemic and epidemic components used 3-month lagged temperature, precipitation, and cumulative incidence as covariates. The spatial spread component assumed the correlation between neighbouring districts followed a power-law structure. Each of the components was modelled using a random effect, which allowed the effects to vary by district. These models were fitted using the hhh4 function in the *surveillance* package in R [22].

3) <u>Supervised ("Y-aware") principal components regression</u> that models the time series of monthly dengue cases from each district as a function of the lags in all of the districts (summarised with principal components), along with seasonality and lags of the weather variables from the same district. A matrix of lagged, standardized log-incidence was generated, including lags from 3-5 months for all districts. We first fit univariate regression for each of the covariates and scaled the variables based on the effect estimates; this effectively provides more weight to variables that are more relevant to the outcome. We then performed principal components analysis (PCA) analysis [23,24] using these rescaled covariates. Based on the variance plots, the top 10 principal components were used in a regression model along with harmonic terms to capture seasonality, an autoregressive correlated random intercept, and lagged temperature and rainfall. These models were fit separately for each district using the *INLA* package in R.

See Supplemental Methods for further details of the three modelling approaches.

*Time series cross validation and model selection*

Considering different covariate combinations and random effect structures, performance of 72 models was evaluated using time series cross validation, including 60 hierarchical Bayesian spatiotemporal models, 9 hhh4 models, and 3 PCA models, across 60 forecast periods (i.e., months) from January 2012 to December 2016.

Each of these 72 models was evaluated using time series cross validation by moving forward the end of the training period by one month at a time. Several forecasting performance measures were applied to evaluate the forecasting performance for a forecast horizon of 1 to 3 months. The Continuous Ranked Probability Score (CRPS), bias, and diffuseness were used to assess the accuracy of forecasting dengue incidence [18], and the Brier score was used to assess the accuracy of forecasting dengue outbreaks [25]. The lower the CRPS, the more accurate the forecast, with a value of 0 indicating a perfect forecast; there is no upper bound to CRPS. Bias values can range from -1 to 1, with a score of zero reflecting no bias. The Brier score can range from 0 to 1, with a lower value indicating more accurate outbreak classification. We evaluated diffuseness (i.e., the width of the predictive distribution) for the top five models with the best performance on the other three metrics. Diffuseness scores closer to 0 are desirable, indicating a narrow (i.e.., sharp) distribution, with poor forecasts having higher diffuseness scores. We used the *scoringutils* package in R to calculate these four metrics [26].

Based on model performance using this time series cross validation approach, we combined a subset of the best models into an ensemble forecast. Models were selected based on a few criteria. First, if multiple models performed similarly in terms of CRPS (the heat map by district and by month can be found in Fig S2 and Fig S3, respectively), we preferred the model with fewer covariates and simpler random effects structures. Second, we evaluated how the models performed district by district and identified several clusters of models that performed better for different sets of districts. We ensured that we had representatives of each of these clusters in the ensemble. Finally, to include a diversity of modelling approaches, we ensured that at least one model from each of the three approaches (Bayesian spatiotemporal, hhh4, PCA). This process resulted in five models that were included in the final ensemble. The weights for the five models were determined based on the CRPS scores and did not vary over time or between districts.

To create the ensemble model, we first generating 10,000 samples derived from the predicted distribution of dengue cases for each of the five models included in the final ensemble. We then calculated the CRPS for each of the five models for the time series cross validation period by comparing the sample predictions against observed dengue cases. Next, we determined weights for each model using (1/CRPS$^2$)/sum(1/CRPS$^2$). Samples then were taken from the predictive distributions of each of the five individual models, with the sampling proportional to the weights. The 2.5th, 50th, and 97.5th percentiles of the resulting distribution were calculated [18].

*Reference model*

A basic reference model was developed to estimate the average monthly dengue incidence for each district without covariates, lagged incidence, or spatiotemporal correlation. Let $Y_{i,t}$ represent dengue cases in district $i = 1,2,..,n$ at time $t = 1,2,..,T$ where $n$ is the total number of districts, and $T$ the time steps (months). The model assumes a Poisson likelihood for observed cases and is defined as:

$$Y_{i,t} \sim Poisson(\mu_{i,t})$$
$$log\left(\frac{\mu_{i,t}}{(p_{i,a[t]})}\right) = \alpha + u_i + \eta_{i,m[t]}$$

$p_{i,a[t]}$ is the population of district $i$ during year $a[t]$, included as an offset to adjust case counts by population. The model incorporates a global intercept $\alpha$, $u_i$ assumes that the mean number of monthly cases across different districts is random and uncorrelated. Additionally, the model includes $\eta_{i,m[t]}$, a cyclic random walk for the calendar month to ensure that seasonal trends are captured while allowing for variations in these trends by district. In line with an earlier study [18], we used the Continuous Ranked Probability Skill Score (CRPSS) to quantify the improvement of a model's predictive performance relative to this reference model. It is defined as

$$CRPSS = 1 - \frac{CRPS_{model}}{CRPS_{reference}}$$

A positive CRPSS indicates better performance than the reference, with a CRPSS of 1 indicating a perfect model. A CRPSS close to zero suggests no improvement, and a negative CRPSS suggests that the model underperforms relative to the reference.

*Availability of code and data*

All code used in the analysis was written in R version 4.2.2 and is available at Github.com/e-dengue/dengue_District_HPC

RESULTS

*Descriptive statistics*

The MDR exhibits notable seasonality in temperature, humidity, and precipitation, with a dry season from December to April and a rainy season from May to November (Fig S4). From 2004 to 2022, there were 648,219 cases of dengue reported in the MDR. Incidence was seasonal and tended to be highest during the rainy season from June to November (Fig 1). Incidence varied substantially among years, from a high of 45,384 cases in 2007 to a low of 5,866 cases in 2014. Tan An was the district that experienced the highest average annual incidence rate at 472 cases/100,000 population, while Go Quao had the lowest average annual incidence rate at 54 cases/100,000 population. The maximum number of dengue cases was recorded in August 2022 (n = 7572), while the minimum number were recorded in December (n = 286) 2021.

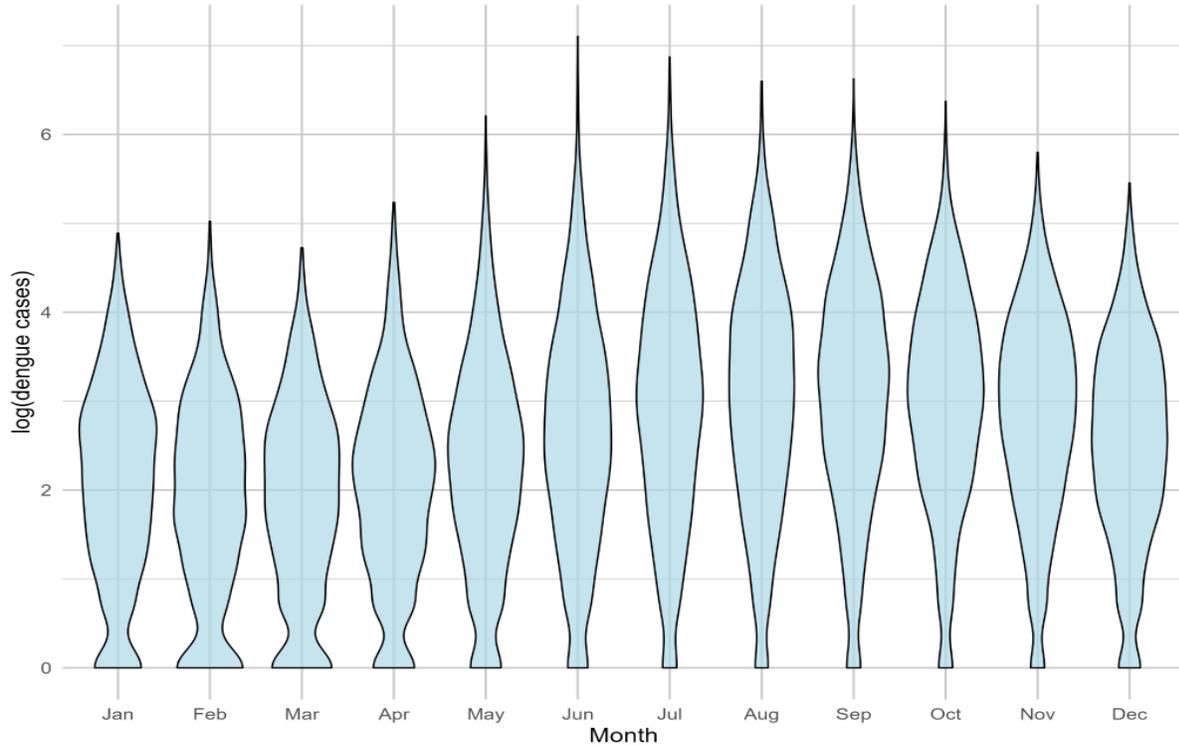

*Fig 1: Monthly dengue incidence (log scale) from 2004 to 2022. The shaded area reflects the density of cases, with wider sections indicating higher density. Case variability was highest from June to October, aligning with the rainy season and increased transmission.*

### *Fit of the models*

The performance metrics for all 72 models tested and for the baseline model are presented in Table S4. We developed an ensemble model that incorporated five of the top-performing individual models, including three hierarchical Bayesian spatiotemporal models, one hhh4 model, and one PCA model (Table 1). The ensemble model demonstrated a lower CRPS (3.2) than any of the 72 individual models, indicating more accurate probabilistic forecasts compared to any individual model. Among the 72 individual models, spatio-temporal model 3 and spatio-temporal model 2 exhibited the lowest CRPS, with values of 3.38 and 3.99, respectively. In contrast, the seasonal reference model had the highest CRPS of 12.2 compared to all 72 individual models, reflecting the lowest predictive accuracy. Among the five models included in the ensemble, the hhh4 model showed the lowest bias at 0.13, followed by Spatio-temporal 3 with a bias of 0.23. Although the ensemble model had l bias compared to the baseline model, its bias metric was not superior to that of its five component individual models.

*Table 1: Summary of model performance metrics and features for dengue incidence prediction three months in advance for the baseline model, the ensemble model, and the five individual models included in the ensemble, aggregated across the Mekong Delta Region districts, 2012-2016. Lower CRPS indicates better accuracy; while lower bias suggests closer alignment with observed values; and lower diffuseness*

*reflects more precise predictions. Weight is the weight each of the five individual models contributed to the ensemble.*

| Model | CRPS | Bias | Diffuseness | Weight | Model features |
|---|---|---|---|---|---|
| Reference | 12·2 | 0·62 | 0·25 | | district effects and seasonality |
| Ensemble | 3·20 | 0·27 | 0·81 | | All those included in the component models below |
| Spatio-temporal model 1 | 4·41 | 0·35 | 0·74 | 0·17 | • 3-month lagged cases<br>• Temporal smoothing with AR(1)* random effects shared across districts<br>• District-level random intercepts (uncorrelated)<br>• Seasonality |
| Spatial-temporal model 2 | 3·99 | 0·30 | 0·67 | 0·20 | • 3-month lagged cases<br>• 3-month lagged average minimum daily temperature<br>• 3-months lagged cumulative precipitation<br>• District-level random intercept with Besag spatial smoothing<br>• Temporal smoothing with AR (1)* random effects (shared across districts)<br>• Seasonality |
| Spatio-temporal model 3 | 3·38 | 0·23 | 0·68 | 0·26 | • 3-month lagged cases<br>• Cumulative incidence over the previous 12, 24, and 36 months (log-transformed)<br>• 3-month lagged average minimum daily temperature<br>• 3-months lagged cumulative precipitation<br>• District-level random intercept with Besag spatial smoothing<br>• Temporal smoothing with AR (1)* random effects (shared across districts)<br>• Seasonality |
| hhh4 model | 4·31 | 0·13 | 0·74 | 0·20 | • 1-month lagged cases within a given district and in neighboring districts<br>• 3-month lagged average daily temperature<br>• 3-monoth lagged cumulative precipitation<br>• Spatial effects modeled using a power-law decay function |

| | | | | | - District-level random intercept<br>- Seasonality |
|---|---|---|---|---|---|
| PCA model | 4·45 | 0·29 | 0·73 | 0·17 | - Lagged cases from all districts used as predictors for other regions<br>- Temporal smoothing with AR (1)* random effects over time<br>- Seasonality captured through sine and cosine components (12-month periodicity) |

AR(1)* : stands for 'Autoregressive of order 1,' meaning the value of a variable at a given time point depends on its value at the previous time point. In our model, AR(1) random effects capture this temporal dependency, ensuring that outcomes like dengue cases are influenced by those at the preceding time point

Besage: spatial smoothing is a method used to model spatial relationships. It assumes that areas close to each other are more likely to have similar outcomes, like disease rates, compared to areas farther apart. Using Besag spatial smoothing means that the risks influence the dengue risk in one district in neighboring districts. If one district has a high number of cases, neighboring districts are likely also to show higher risks

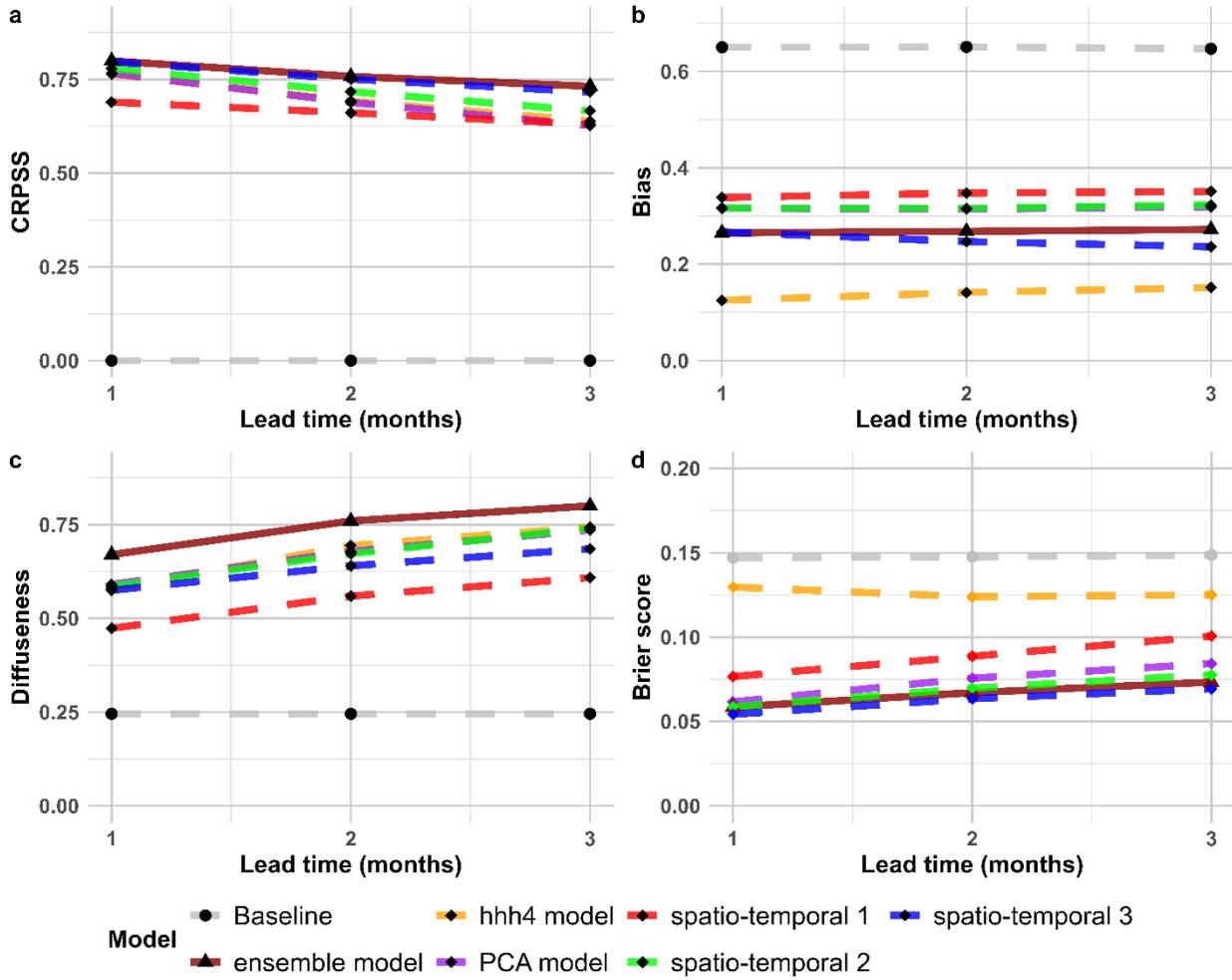

*Fig 2: Comparison of performance metrics for the baseline (reference model), the ensemble model, and the five individual models included in the ensemble across forecast horizons of 1 to 3 months, aggregated across the Mekong Delta Region districts, 2012-2016. Panel (a) shows CRPSS, with higher values indicating a better forecast compared to the baseline model. Panel (b) displays bias, with values closer to zero reflecting more accurate predictions. Panel (c) presents diffuseness, which measures the spread of the predicted probabilities, with lower values indicating greater precision.*

The performance of all models tended to decline as the forecast horizon increased from 1 to 3 months, with CRPSS decreasing and diffuseness increasing with longer lead times, although bias tended to be consistent across time horizons (Fig 2). The decreasing predictive performance of the ensemble model with increasing time horizon is further illustrated in Fig 3, which shows the ensemble forecast mean and its 95% predictive interval compared to the observed dengue cases. For example, during the 2012 dengue wave, the 3-month-ahead prediction (blue line) underestimated the magnitude of the outbreak, whereas as we moved closer to the actual date (2-month and 1-month predictions), the forecasts became more accurate, with the 1-month-ahead prediction (orange solid line) tracking closest to the observed cases and the uncertainty intervals narrowing. Nevertheless, despite the lower accuracy of the 3-month predictions, these longer-term forecasts offer early indications of rising cases well in advance of the actual outbreak, giving public health officials time to prepare. For instance, for the period June through September 2012, the 3-month horizon prediction starts showing the upward trend in dengue cases well before the outbreak fully occurs (Fig S4).

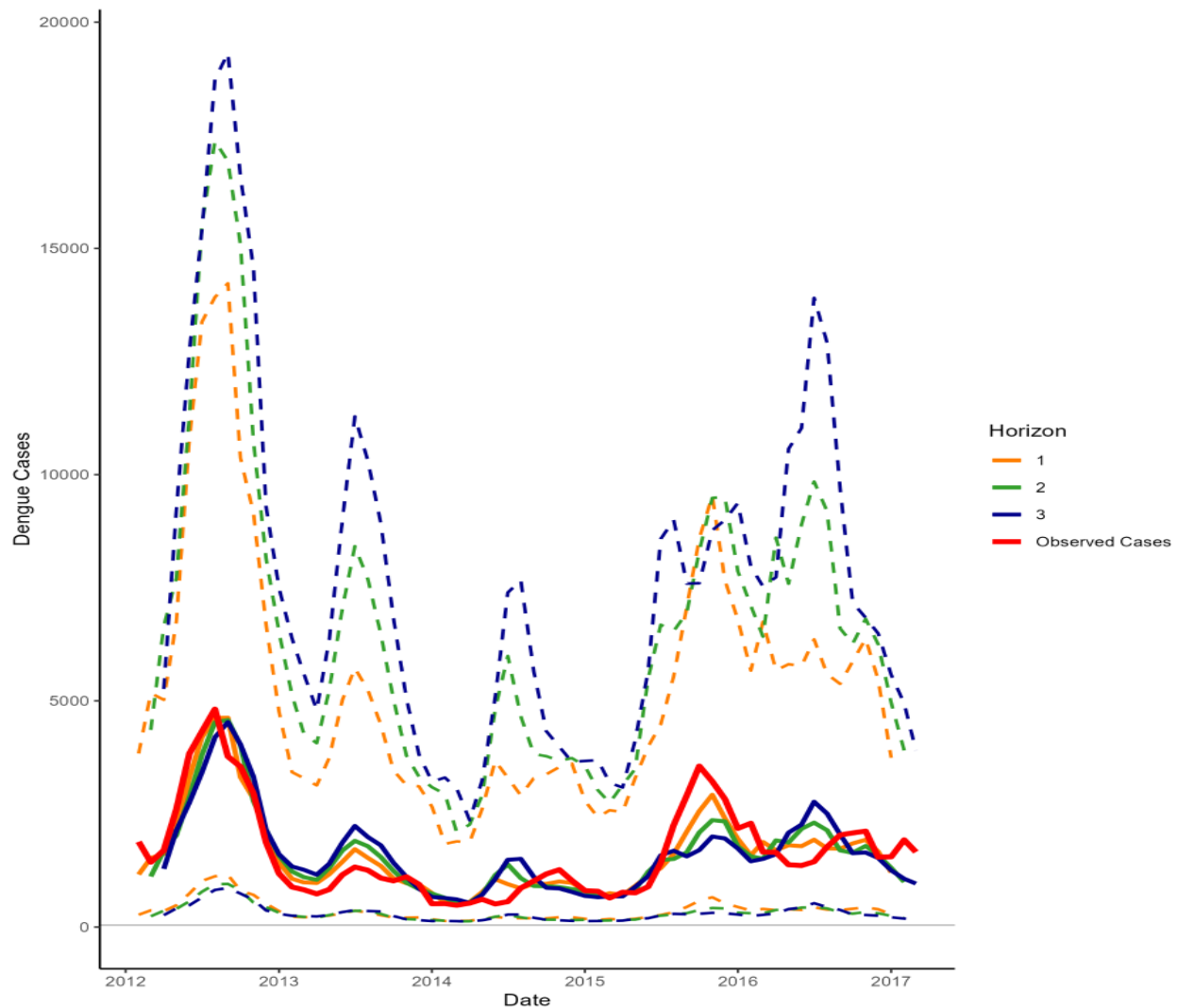

*Fig 3 : Observed vs. forecasted dengue cases (mean and 95% predictive interval) using the ensemble model, by time horizon, aggregated across the Mekong Delta Region districts, 2012-2016. The solid red line represents observed cases, the solid blue line forecasted cases with a 3-month horizon, the solid green line forecasted cases with a 2-month horizon, and the solid yellow line forecasted cases with a 1-month horizon. The dashed lines correspond to the 95% credible intervals for the forecasts.*

The ensemble model's ability to classify outbreak and non-outbreak periods was evaluated using the Brier score and four different outbreak threshold definitions -- mean + 2 standard deviations, 95th percentile, Poisson threshold, and fixed thresholds based on incidence rates, as described in the Methods. Each threshold involves trade-offs, and the choice of which to use should align with the surveillance system's programmatic goals. For example, a high fixed threshold will have low sensitivity, meaning fewer outbreaks are detected. However, this also reduces the likelihood of false positives. The highest classification accuracy was achieved using a stringent outbreak threshold set at 50 cases per 100,000 population, followed by the 95th percentile threshold, and then the mean plus two standard deviations threshold (Fig 4). The accuracy was generally higher in the summer months when case counts were higher.

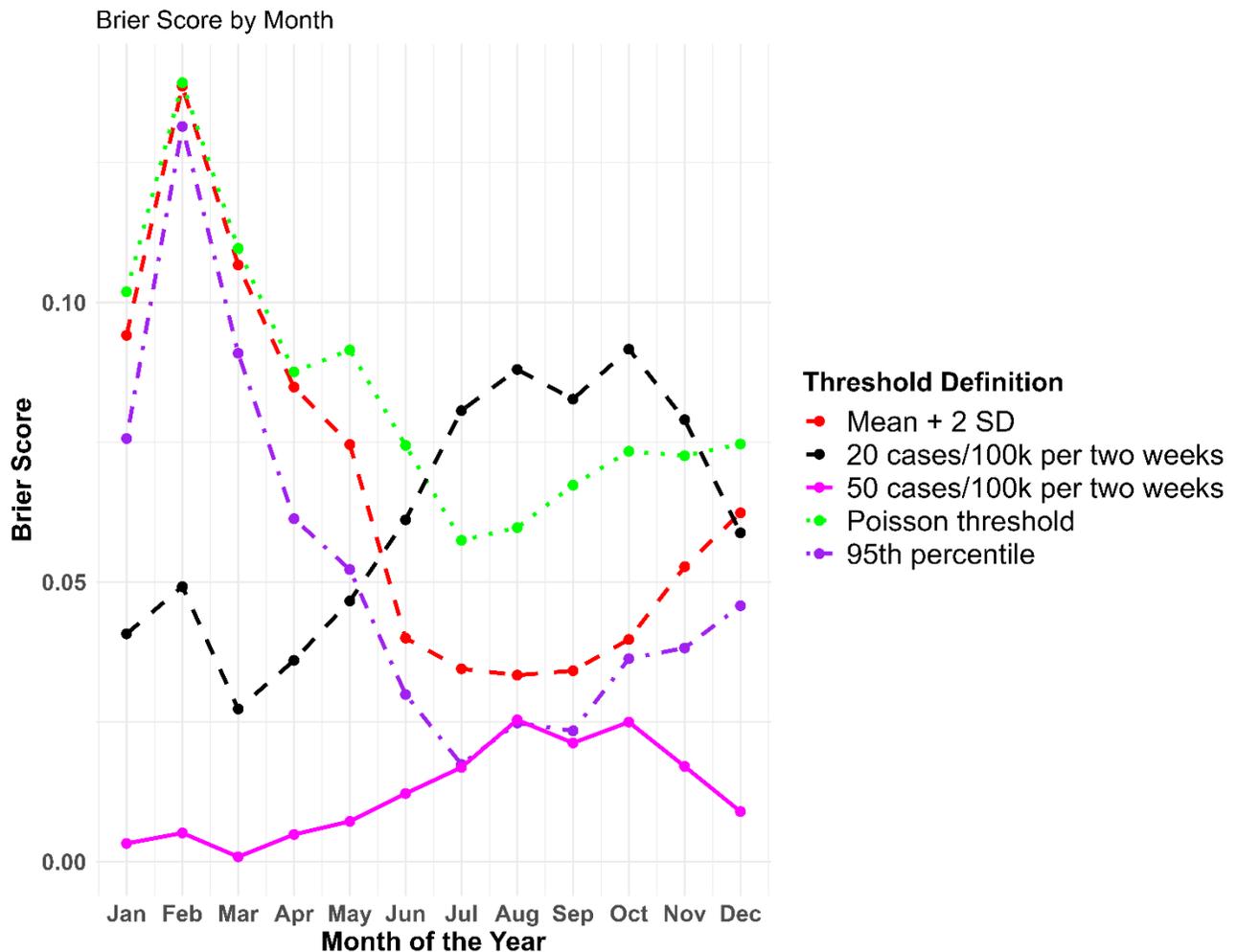

*Fig 4: Brier score averaged by month across the three time horizons using the ensemble model with five different threshold definitions, aggregated across the Mekong Delta Region districts, 2012-2016. The Brier score can range from 0 to 1, with 0 representing the optimal value.*

In addition, to using the Brier score, the accuracy of the ensemble model for classifying outbreaks can be evaluated by comparing the predicted probability of an outbreak and how frequently an outbreak actually occurred for different bins of predicted probabilities. If the points align along the diagonal, this indicates a well-calibrated model (Fig S5). In general, at lower predicted probabilities, the model underestimated the risk of outbreaks, while at higher predicted probabilities, the model overestimated the risk. The accuracy differed based on the threshold used, with the fixed thresholds generally exhibiting better calibration.

*Model performance out-of-sample*

We evaluated the ensemble model's performance, along with that of its five individual model components and the baseline model, during the 2017-2022 out-of-sample evaluation period (Fig 5). We found lower classification accuracy (higher Brier score) than during the cross-validation period, but still with reasonable accuracy for 1-, 2-, and 3-month horizons. The relative performance of the models across all metrics was similar to what was seen during the cross-validation period, with the lowest Brier score obtained from the ensemble model. As an example of the ensemble model's out-of-sample prediction performance, Fig S6 shows that the model captured the overall seasonal dynamics and anticipated a sharp rise in cases during May-August 2022. The ensemble model's performance over time, as shown in (Fig 6, and Fig S7) demonstrates its ability to track dengue trends; however, misalignments are evident in the years 2019 and 2022.

In 2022, during the post-COVID recovery phase, the ensemble model struggled to predict dengue outbreaks accurately. As seen in Fig 6, sharp increases in observed cases during certain periods of 2022 were not matched by the model's predictions, or the model predictions only increased in step with the observed cases. As a result of these misalignments, the accuracy of predicting epidemics was compromised, with notable variations observed across different thresholds and years. For instance, the overall accuracy fluctuated considerably between years, achieving the highest performance in 2017 (77%) and 2018 (73%), followed by a decline in 2019 (65%). Although a slight recovery was noted in 2020 (72%) and 2021 (71%), the accuracy dropped again in 2022 (56%). These inconsistencies highlight the ongoing challenges in balancing the detection of outbreaks and minimising false alarms, particularly when applying the model to distinct groups of years with varying outbreak dynamics (Table S3).

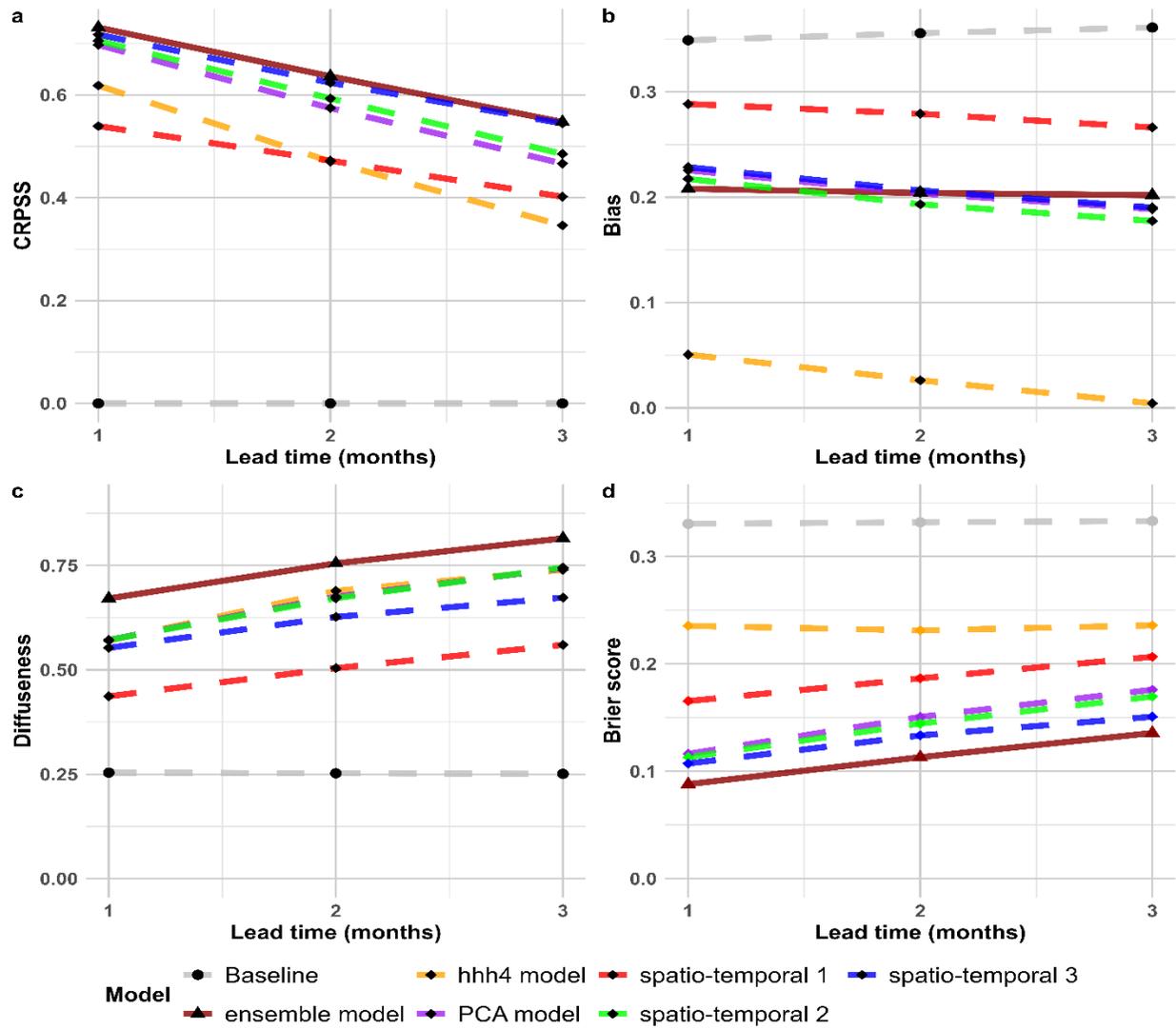

*Fig 5: Performance metrics for the baseline (Reference) model, the ensemble model, and the five individual models included in the ensemble across forecast horizons of 1 to 3 months, Mekong Delta Region, during the 2017-2022 out-of-sample evaluation period. Panel (a) shows CRPSS, with higher values indicating a better forecast compared to the baseline model. Panel (b) displays bias, with values closer to zero reflecting more accurate predictions. Panel (c) presents diffuseness, which measures the spread of the predicted probabilities, with lower values indicating greater precision. Panel (d) represents the Brier score, using Mean+2 SD, which can range from 0 to 1, with 0 representing the optimal value.*

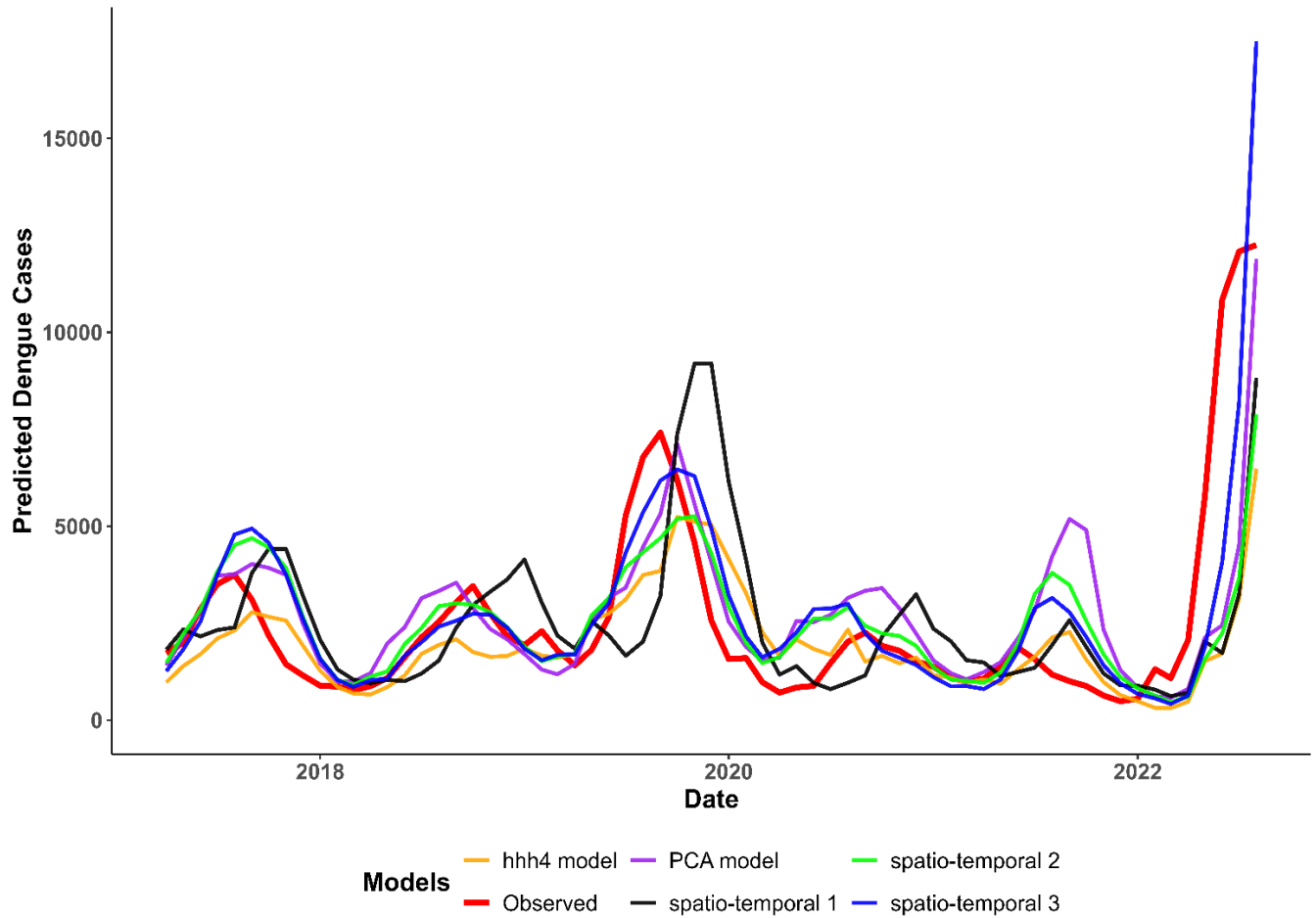

*Fig 6. Time-series plot of predicted dengue cases from various models compared to observed cases (2017-2022), showing the performance of the top five individual models. This plot illustrates the temporal patterns and predictive performance of different models over the years. Notably, there is a visible misalignment in predictions for the years 2019 and 2022, indicating potential challenges in model accuracy during these periods.*

DISCUSSION

In this study, we have demonstrated that an ensemble of probabilistic models can predict dengue outbreaks at 1-3-month lead times, specifically at the district level in the MDR region of Vietnam. By combining individual spatial-temporal, semi-mechanistic (hhh4), and supervised PCA models, our ensemble outperformed individual models. The ensemble model's strength lies in its ability to integrate key features of the five individual models included in the ensemble, such as temporal smoothing through autoregressive correlation, district-level random effects, meteorological factors (i.e., temperature and precipitation), and lagged cases from various periods. The ability to generate probabilistic predictions can enhance decision-making for proactive public health interventions, providing timely early warnings for targeted dengue control measures [28]. Disruptions to historical seasonal patterns in recent years proved a challenge for the accuracy of the models, and further monitoring is needed to determine if the model performance improves again as epidemic patterns settle back into a more typical seasonality.

The performance of the ensemble model varied across geographic locations, forecast horizons, and months of the year. Our analysis demonstrated the model's accuracy in predicting spatiotemporal variations in dengue cases and the occurrence of outbreaks with forecast horizons of up to 3 months. However, as lead time increased, the model's predictive performance deteriorated, and the associated uncertainty grew. This trend aligns with other findings from Vietnam and from similar studies in other settings and diseases, where longer forecast horizons tend to result in reduced accuracy due to increasing variability and complexity in the underlying dynamics [29]. Outbreaks are difficult to predict, as assessed using scoring rules across various types of outbreak thresholds [30].

The findings of this study have direct implications for improving dengue prevention and control efforts in the Mekong Delta. The ability to forecast dengue outbreaks at the district level enables local public health officials to take timely action to reduce the effects of dengue outbreaks, including proactive vector-control interventions such as larval control, insecticide spraying, or community mobilization. The use of probabilistic forecasting provides decision-makers with credible intervals around potential outbreak scenarios, enabling them to plan interventions based on risk variability. For instance, if an area shows a high probability of an outbreak, preventive measures can immediately be applied, while regions with lower risks can adopt a more monitored approach, optimizing the use of limited resources.

The district-level focus of the model is particularly valuable for Vietnam's health system. The model's ability to provide fine-scale forecasts ensures that local health departments can tailor their interventions based on the specific epidemiological and environmental context of each district, rather than relying on broader, less targeted provincial-level forecasts. In addition to supporting proactive interventions, the model aligns well with Vietnam's efforts to adapt to climate change, which is a recognized driver of vector-borne diseases like dengue. The government is increasingly seeking tools that can incorporate climate predictions into public health planning. Our model, which leverages seasonal climate data, can serve as a component of an early warning system that links environmental data with health outcomes.

A previous study by Colón-González et al. [18], generated seasonal dengue forecasts at the provincial level in Vietnam, while our approach offers more localised forecasts, which are crucial for on-the-ground interventions. Moreover, our ensemble incorporated three types of modelling approaches, which provides more robust predictions across a range of different districts.

Despite its strengths, this study also has limitations. Our modelling framework integrates key factors influencing dengue incidence, including meteorological variables. However, it does not directly account for other determinants of disease incidence, such as variations in serotypes, immunity, and human mobility. In addition, the Covid-19 epidemic could have affected dengue reporting during 2020-2022. In our analysis, spatial and spatiotemporal random effects are incorporated in each model within the ensemble to account for unmeasured variability associated with these factors. While we were able to forecast at the district level, scaling down to even finer resolutions, such as the commune (the third-level administrative unit below the province and district), would be beneficial for planning public health responses but remains a challenge due to the sparsity and inconsistency of data at these levels. We also found that different epidemic thresholds yielded varying results. This variability highlights the sensitivity of outbreak detection to the choice of threshold. Adjusting thresholds based on historical data, such as previous outbreak patterns, and expert knowledge of local epidemiological trends could help improve the model's accuracy in identifying true outbreaks while minimizing false alarms. For example, thresholds could be dynamically set to reflect seasonal trends or differences between high- and low-transmission years, allowing for more context-specific detection.

Another limitation to this study is, that the observed misalignments in 2019 and 2022 underscore the ensemble model's ability to adapt to rapidly changing conditions. During the post-pandemic recovery in 2022, changes in mobility patterns, environmental conditions, and transmission dynamics led to underpredicted outbreaks. These challenges highlight the need for greater flexibility in the ensemble approach. Introducing dynamic weighting [27], where the influence of component models adjusts based on real-time data and emerging trends, could improve the model's responsiveness.

CONCLUSION

This study shows that using an ensemble of models can effectively forecast dengue outbreaks at the district level in the Mekong Delta Region with lead times of one to three months. However, its performance during atypical years, such as 2019 and 2022, revealed limitations in adapting to systemic disruptions and evolving transmission dynamics, underscoring the need for more flexible approaches. Incorporating dynamic weighting, which adjusts model contributions in real time based on changing conditions, could enhance the model's adaptability and robustness. The ensemble's district-level focus is especially important for local interventions, and we plan to incorporate the ensemble into software for an early warning system that will integrate real-time meteorologic and dengue surveillance data to be used by districts to head-off dengue outbreaks through proactive interventions. The ensemble model development approach we used can also be adapted for use in other countries and for predicting other diseases like malaria or Zika. While the model works well at the district level, further improvements are needed for more detailed predictions at the commune level. Incorporating additional data, such as dengue serotypes and human movement patterns, could make the system even more accurate.

*Conflict of Interest*

The authors confirm that there are no conflicts of interest related to this work.

**Appendix**

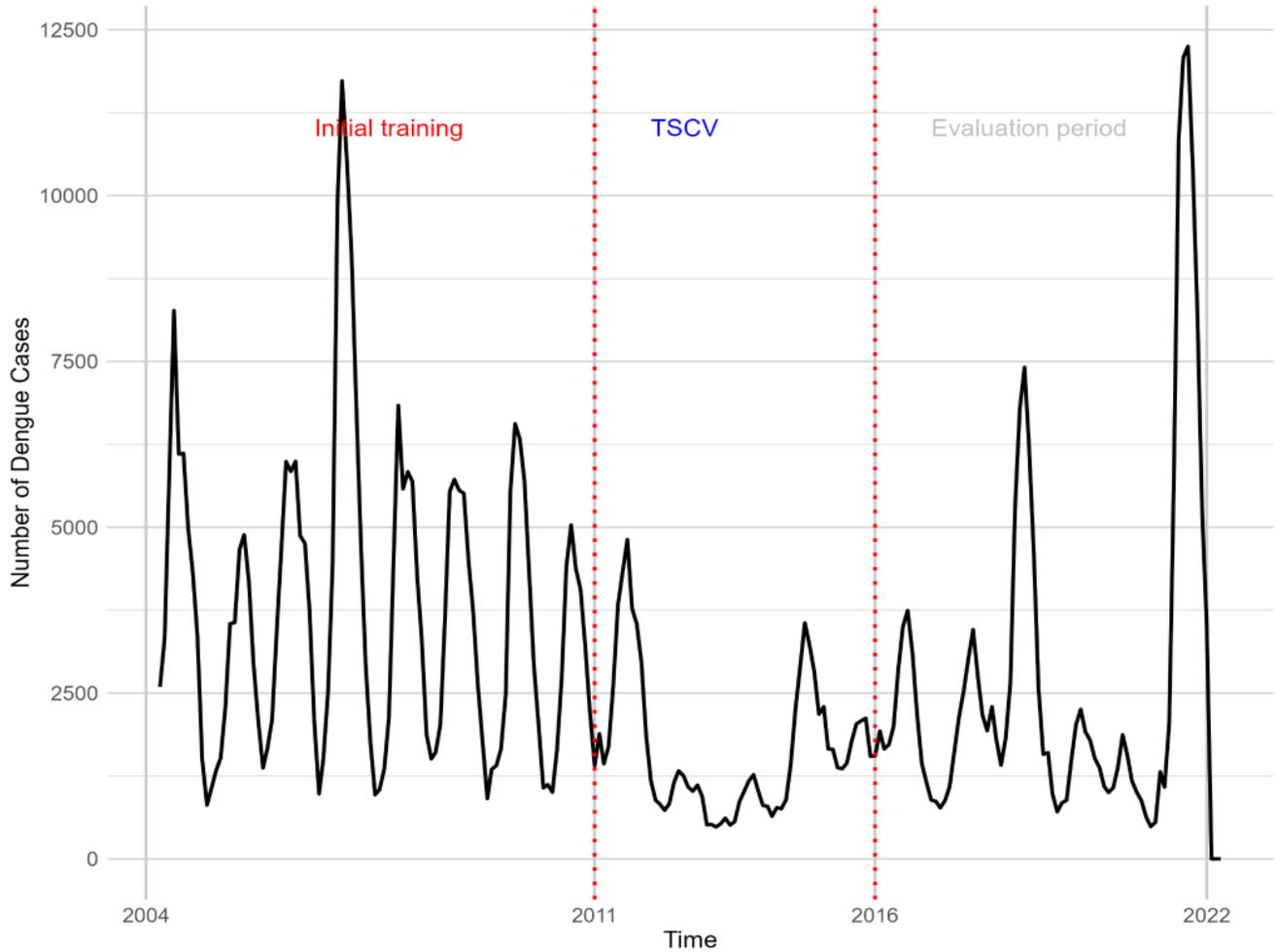

*Fig S1: time series of dengue cases from 2004 to 2022, highlighting the initial training period (red), the time-series cross validation (TSCV) phase (blue), and the evaluation period (grey). The initial training data is used for model calibration, followed by TSCV for model validation, and concluding with an evaluation period for assessing forecast accuracy.*

*Table S1: Socio-economic dataset (data source: General Statistics Office of Vietnam)*

| Name | Unit | Frequency | Level of data |
|---|---|---|---|
| Population density | $Km^2$ | Yearly | District |
| Population data | pop | Yearly | District |

| In-migration rate | % | Yearly | Province |
|---|---|---|---|
| Poverty rate | % | Yearly | Province |
| Hygienic water access | % | Yearly | Province |
| Hygienic toilet access | % | Yearly | Province |
| Monthly average income | VND[a] | Yearly | Province |
| Monthly average income per capita | VND[a] | Yearly | Province |
| Total passenger by province each year | Million-person time | Yearly | Province |
| Urbanization rate | % | Yearly | Province |
| Specially land use (includes land used by the government offices; public services construction facilities; security and national defence land; land for non-agricultural production and business, and public land) | $m^2$ | Yearly | Province |

[a]Vietnamese dong

Table S2: Preventive measures/entomologic indices dataset (data source: Vietnam National Surveillance System)

| Name | Unit | Frequency | Level of data |
|---|---|---|---|
| Larvae index (BI[)]) | No unit | Monthly | District |
| CI[b] larvae | % | Monthly | District |
| HI[c] larvae | % | Monthly | District |
| Mosquito index (DI[d]) | % | Monthly | District |
| HI[c] mosquitoes | % | Monthly | District |
| Breeding site elimination campaigns | Interventions (N) | Monthly | District |
| Active spraying | Intervention (N) | Monthly | District |
| Large scale spraying for epidemic response | Intervention (N) | Monthly | District |
| Communication and/or training | Intervention (N) | Monthly | District |

| Name | Unit | Frequency | Level of data |
|---|---|---|---|
| Number of outbreaks detected | Outbreaks | Monthly | District |
| Number of outbreak responses | Outbreaks | Monthly | District |

[a]Breteau index; [b]Container index; [c]House index; [d]Density index

*Table S3: The tested models using different covariates (Full excel sheet also provided)*

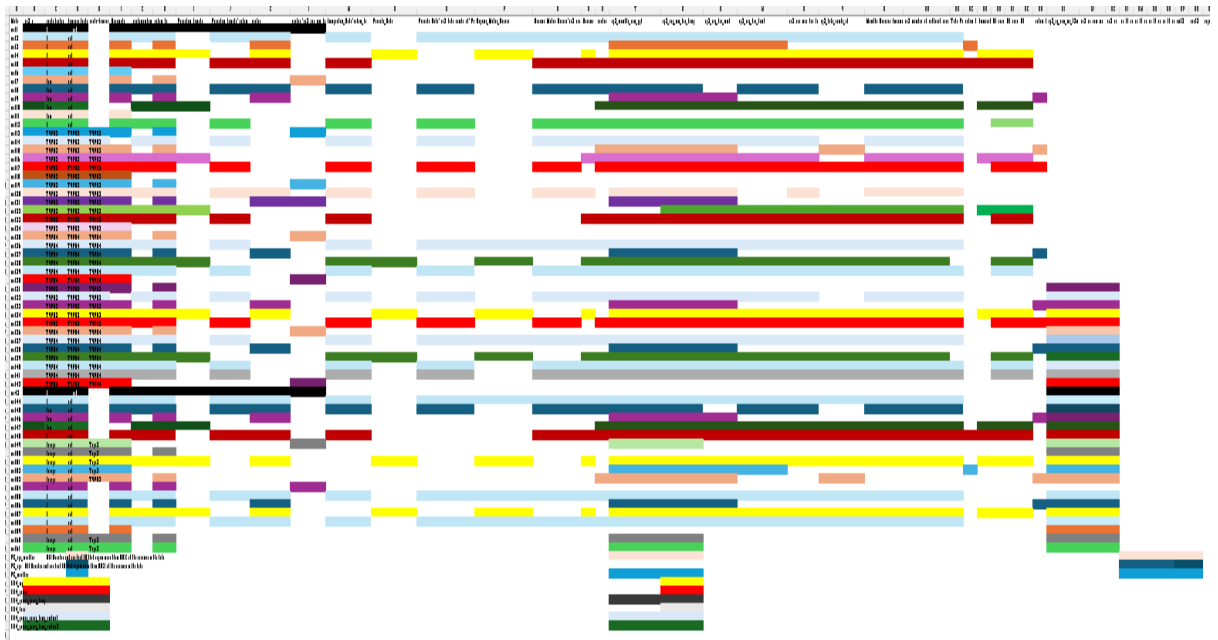

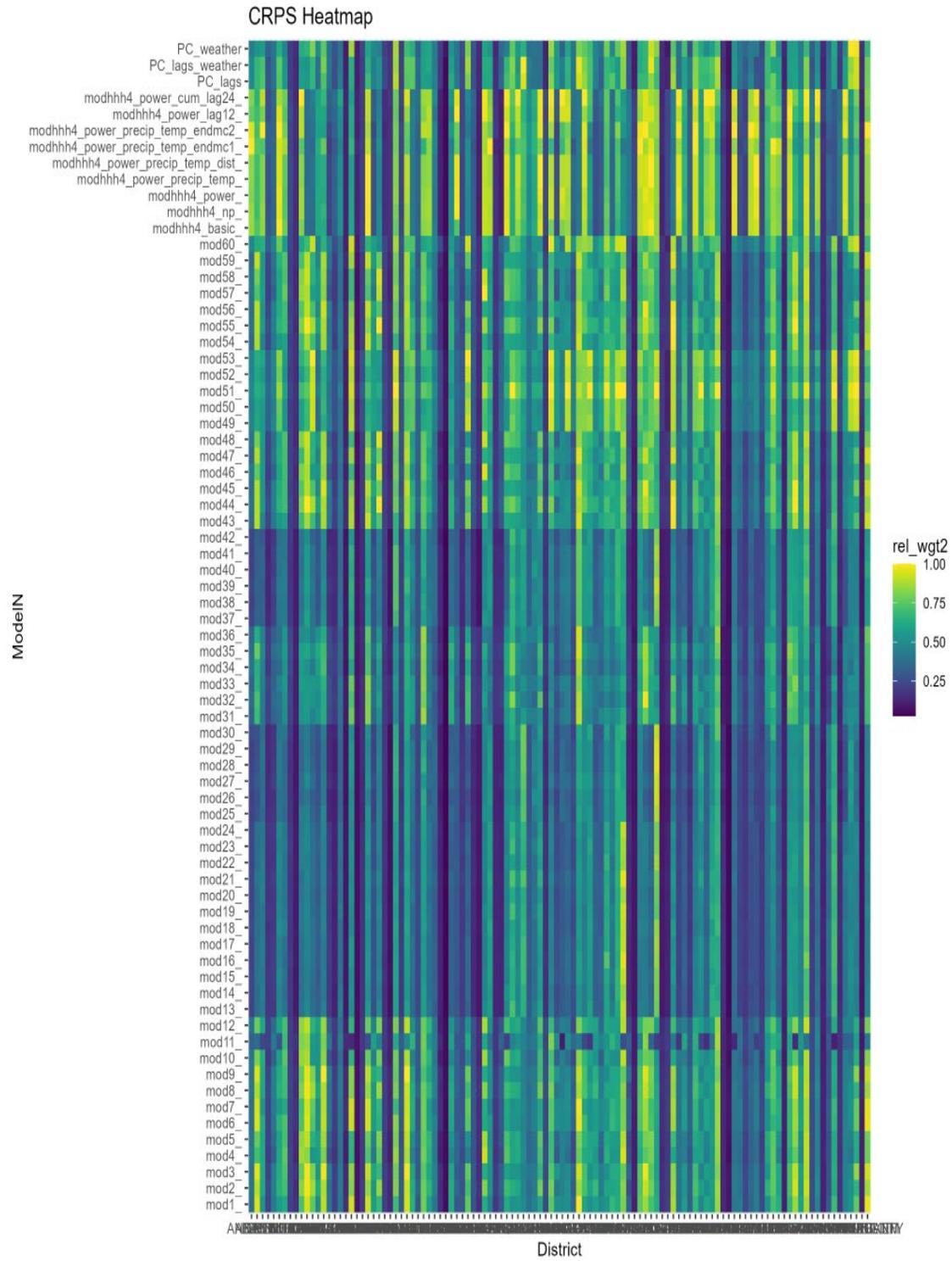

*Fig S2: Heat map of CRPS relative values, with the x-axis representing districts and the y-axis showing the tested models.*

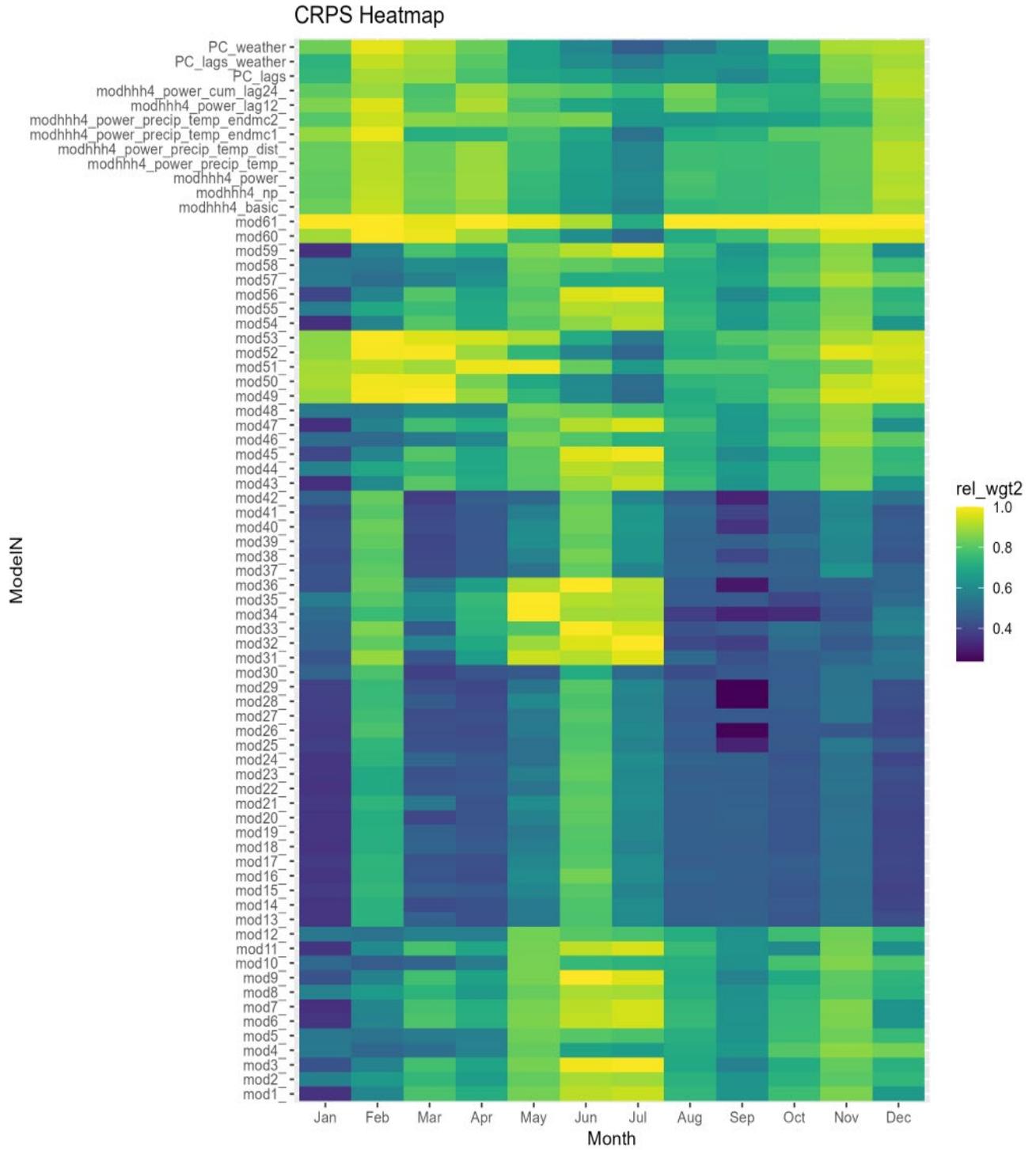

*Fig S3: Heat map of CRPS relative values, with the x-axis representing months and the y-axis showing the tested models.*

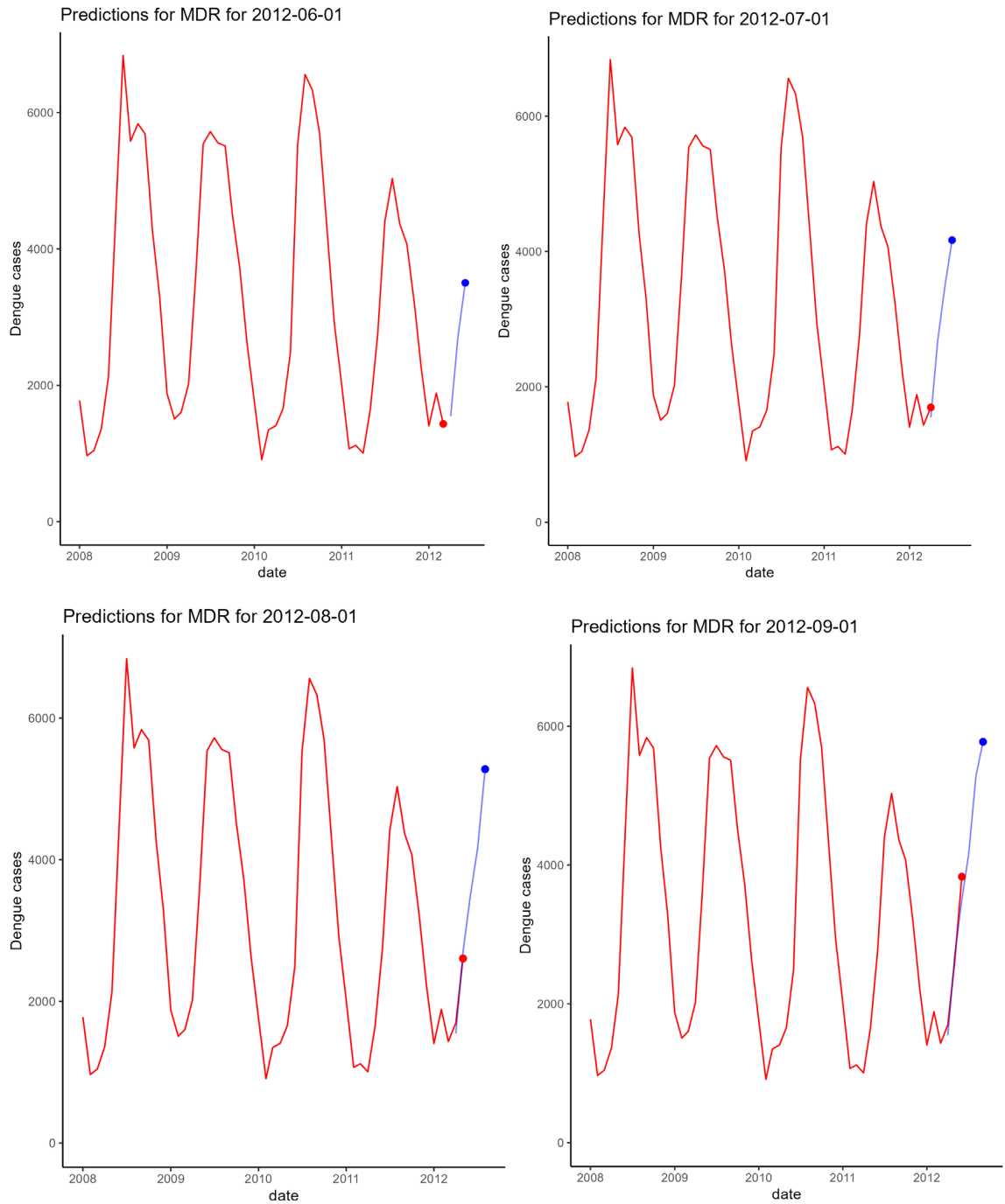

*Fig S4: 3-month ahead predictions of dengue cases for the Mekong Delta Region using the ensemble model. The red lines represent observed dengue cases, while the blue lines denote predicted cases. The forecasts demonstrate the model's ability to capture seasonal trends and predict future outbreaks.*

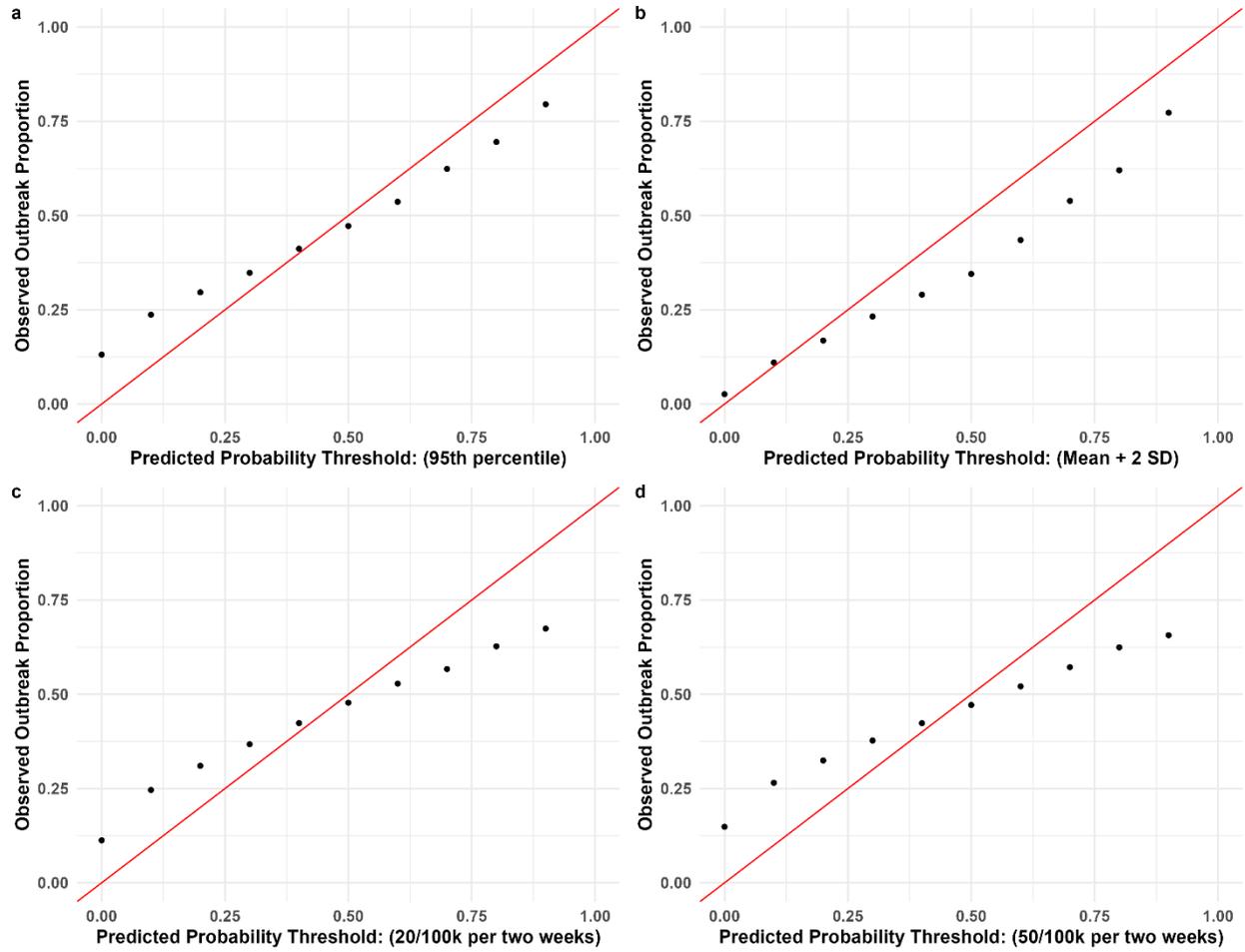

*Fig S5: Accuracy of the ensemble model for selected outbreak thresholds based on the probability of predicting an outbreak compared to the observed outbreak proportion, Mekong Delta Region, 2012-2016. X-axis represents bins of predicted probabilities (0-<0.1, 0.1-<0.2, 0.2-<0.3…0.9-1). Y-axis represents the proportion of observations for which the observed exceeded the predicted.*

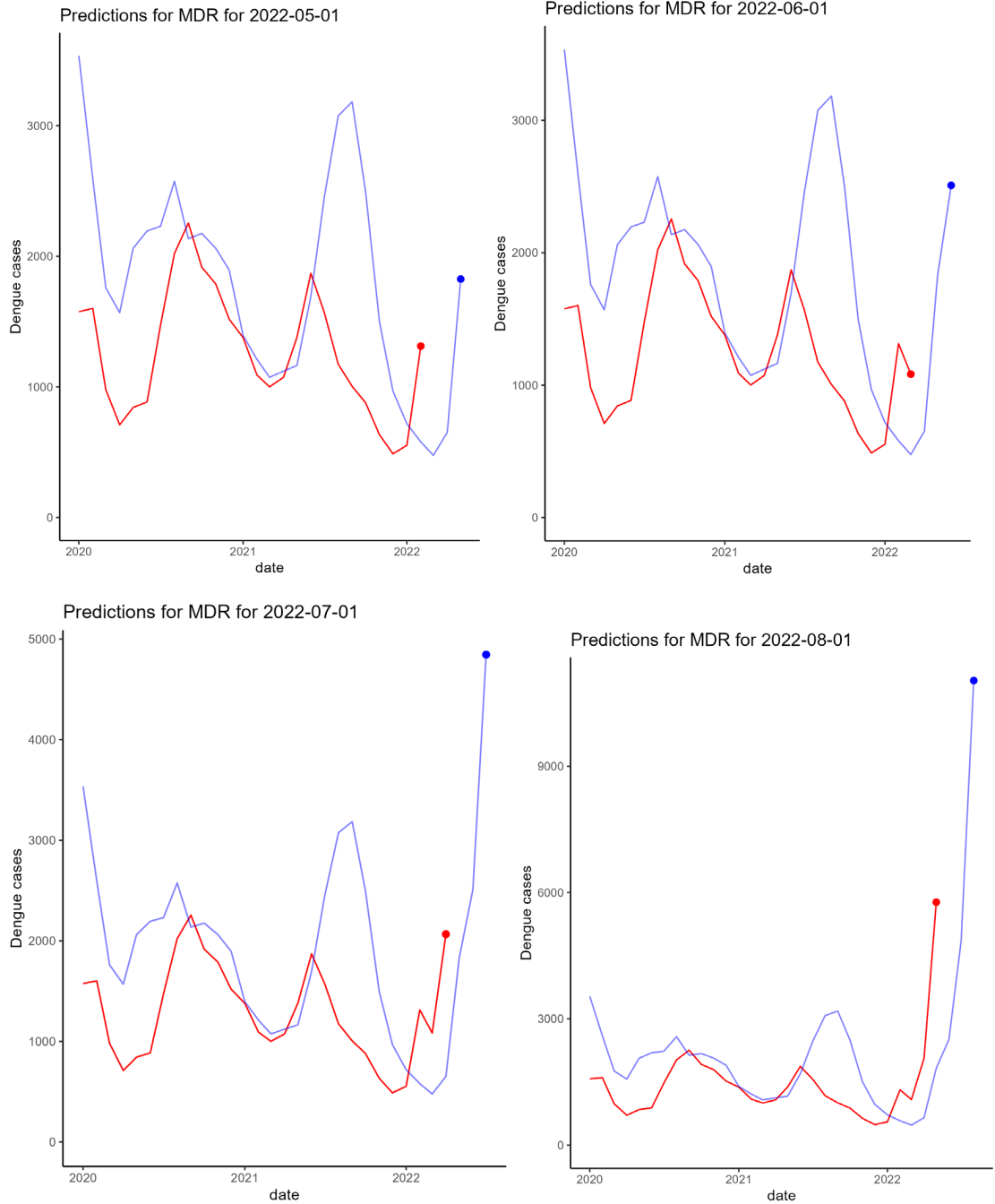

*Fig S6 3-month ahead out-of-sample (2017-2022) predictions of dengue cases using the ensemble model for the Mekong Delta Region. The red lines represent observed dengue cases, while the blue lines denote predicted cases. The forecasts demonstrate the model's ability to capture seasonal trends and predict future outbreaks*

*Table S3: ensemble performance in predicting outbreaks using different thresholds for the two groups years (2017,2018 and 2020) and years (2019 and 2022)*

| Threshold | Sensitivity | Specificity | Positive predictive value (PPV) | Sensitivity | Specificity | Positive predictive value (PPV) |
|---|---|---|---|---|---|---|
| | Years (2017, 2018 and 2020) | | | Years (2019 and 2022) | | |
| 20 cases/100k cases | 25% | 93% | 40% | 22% | 85% | 34% |
| 50 cases/100k cases | 7% | 94% | 55% | 5% | 89% | 49% |
| 95th percentile | 42% | 92% | 51% | 38% | 85% | 47% |
| Mean+2 SD | 57% | 79% | 33% | 49% | 60% | 22% |
| Poisson distribution | 65% | 73% | 34% | 46% | 60% | 18% |

**Supplemental Methods**

**Hierarchical Bayesian spatiotemporal models**

Let $Y_{i,t}$ be the number of dengue cases in the district $i = 1,2,..,n$ and time $t = 1,2,..,T$, where $n$ is the total number of districts in the data set, and $T$ is the total number of time steps for which the model is fitted. The models used a Poisson likelihood when fitting to the number of observed cases. The generic model is defined as:

$$Y_{i,t} \sim Poisson(\mu_{i,t})$$

$$\mu_{i,t} = log\left(\frac{\mu_{i,t}}{(p_{i,a[t]})}\right) = \alpha + log\left((Y_{i,t-3} + 1)/(p_{i,a[t]})\right) + \sum_{k} \beta_k X_{k,i,t} + \eta_{i,m[t]} + \delta_{i,a[t]} + \theta_i$$

Where:
- $Y_{i,t-3}$ is the lagged number of monthly dengue cases in district $i$ and time $t$ lagged 3 months.
- $\alpha$ is the intercept.
- $p_{i,a[t]}$: is the population for district i and year $a[t]$, included as an offset to adjust case counts by population.
- $\beta_k X_{k,i,t}$: represents the fixed effects for the covariates, with $\beta_k$ being the corresponding coefficients.
- $\delta_{i,a[t]}$ is a random effect with a first order autoregressive structure that captures temporal variation. In some models, the random effect is the same across districts, while in other models, there was a separate AR(1) random effect for each district, with hyperparameters (e.g., variance) shared across districts

- $\eta_{i,m[t]}$ : is seasonality which has a cyclic 1st order random walk process for time, with a separate random walk estimated for each district but with a shared hyperparameter on the variance.
- $\theta_i$: is a random effect for each district that accounts for local variability that is unique to the district (unstructured noise), as well as the influence of neighbouring districts (spatial correlation) modelled by some models like Besag-York-Mollie.

Delayed effects for meteorological factors were accounted for by a lag of 3 months. Normal priors were set for the fixed effects with mean 0 and variance 1, while the global intercept had a prior with mean 0 and variance 25. Spatial dependencies were modelled using the Besag model, with a log-gamma prior for precision to ensure moderate smoothness. Penalising Complexity (PC) priors were applied across all random effects. The BYM and BYM2 models incorporated spatial and district-specific random effects, using PC priors to encourage smoother spatial variation. Temporal autocorrelation was modelled using an AR (1) process, with PC priors applied to both the temporal and autocorrelation parameters. For unstructured random effects, PC priors were used to capture district-specific variability. Seasonal trends were modelled using an RW1 process (Random Walk of Order 1 and is a commonly used model in Bayesian statistics for smoothing temporal or spatial data
), with PC priors providing flexibility in the smoothness of monthly effects. Models were fitted in R version 4.2.2 using the *INLA* package.

**hhh4 models**

This model predicts the number of dengue cases in a district $i$ at time $t$ autocorrelation, and spatial spread. Each of these components—endemic, epidemic, and spatial—are modelled as a function of covariates. The outcome $Y_{i,t}$ representing the observed dengue cases follows a Negative Binomial distribution:

$$Y_{i,t} \sim NegBin(\mu_{i,t}, \psi)$$

Where $\mu_{i,t}$ is the mean and $\psi$ controls the dispersion. The model decomposes $\mu_{i,t}$ as:

$$\mu_{i,t} = v_{i,t} + \lambda_{i,t} Y_{i,t-1} + \phi_{i,t} \sum_{j \neq i} w_{j,i} Y_{j,t-1}$$

**Endemic Component** $v_{i,t}$
This term captures seasonality and covariates effect

$$\log(v_{i,t}) = \alpha + \log(p_{i,a[t]}) + \sin\left(\frac{2\pi t}{12}\right) + \cos\left(\frac{2\pi t}{12}\right) + \sum_k \beta_k X_{k,i,t}$$

**epidemic component (autoregressive) Component** $\lambda_{i,t} Y_{i,t-1}$:

This component models the influence of past cases within the same district.

$$log(\lambda_{i,t} Y_{i,t-1}) = \alpha + \sum_k \beta_k X_{k,i,t}$$

**spatial spread component (neighbourhood) Component** $\phi_{i,t} \sum_{j \neq i} w_{j,i} Y_{j,t-1}$

This captures the influence of neighbouring districts' past cases on the current district.

$$log(\phi_{i,t}) = \alpha + \sum_k \beta_k X_{k,i,t}$$

Where $\alpha$ is the intercept, $p_{i,a[t]}$ is the population offset for the district $i$ at time $a[t]$, $\beta_k X_{k,i,t}$ is the fixed covariates (e.g., 3-month lagged temperature and precipitation), $w_{j,i}$ represents the spatial weights, reflecting the influence of neighbouring districts through a power-law function, $\sin\left(\frac{2\pi t}{12}\right)$ and $\cos\left(\frac{2\pi t}{12}\right)$ are the seasonal harmonic terms to capture seasonality over a 12-month cycle.

Models were fitted in R version 4.2.2 using the surveillance package[22].

## Supervised ("Y-aware") principal components regression

For each district $i$ at time $t$, the dengue incidence $Y_{i,t}$ is modelled as a function of multiple covariates and lagged predictors. The covariate matrix $X$ includes standardised lagged covariates for all districts[24].

A univariate linear regression is performed between the log-transformed dengue case counts and each covariate $X_k$ to calculate the regression slope:

$$\log(Y_{i,t}) = \beta_k X_{k,i,t} + \varepsilon_{i,t}$$

Where $\beta_k$ is the regression coefficient for covariate $k$. $X_{k,i,t}$ is the value of covariate $k$ for district $i$ at time $t$, and $\varepsilon_{i,t}$ is the error term accounting for unexplained variability.

Each covariate is then multiplied by its corresponding slope:

$$X'_{k,i,t} = \beta_k X_{k,i,t} - mean(\beta_k X_{k,i,t})$$

After rescaling, principal components analysis is applied to the matrix $X'$ to extract the principal components (PCs). These PCs are new variables that are linear combinations of the original covariates, capturing the directions of maximum variance.

$$PC_p = \sum w_{p,k} X'_{k,i,t}$$

- $PC_p$ is the value of the $p$-th principal component.
- $w_{p,k}$ is the weight (loading) of covariate k in the $p$-th principal component.

The top 10 components, which explain most of the variance, were retained for further analysis.

The final regression model predicts dengue incidence by combining the PCs, seasonal components, and an autoregressive term.

$$\log(\mu_{i,t}) = \alpha + \log(p_{i,a[t]}) + \sum_{p=1}^{P} \beta_p \, PC_p + \sin\left(\frac{2\pi t}{12}\right) + \cos\left(\frac{2\pi t}{12}\right) + \sum_k \beta_k X_{k,i,t} + \delta_{i,a[t]}$$

where:

- $\mu_{i,t}$: is the expected number of cases in the district $i$ at time $t$.
- $\alpha$: is the intercept representing the baseline level of dengue incidence.
- $\log(p_{i,a[t]})$: is the population offset for the district $i$.
- $\sum_{p=1}^{P} \beta_p \, PC_p$: is the sum of the principal components, each weighted by its coefficient $\beta_p$.
- $\sin\left(\frac{2\pi t}{12}\right)$ and $\cos\left(\frac{2\pi t}{12}\right)$ are the seasonal effects using sine and cosine terms.
- $\delta_{i,a[t]}$ is the AR (1) random effect capturing temporal correlations.

Models were fitted in R version 4.2.2 using the INLA package.